\documentclass[twocolumn,twoside,slac_two]{revtex4}
\usepackage{graphicx}
\usepackage{fancyhdr}
\begin{document}

\title{Charm-Production in
        \(e^{+}e^{-}\) Annihilation Around 4~GeV}


\author{B.~W.~Lang}
\affiliation{University of Minnesota, 116 Church St. S.E., Minneapolis, MN 55455}

\begin{abstract} 
% Insert abstract here.
Using the CLEO-c detector at the Cornell Electron Storage Ring, we have measured 
inclusive and exclusive cross sections for the production of $D^+$, $D^0$ and 
$D_s^+$ mesons in $e^{+}e^{-}$ annihilations at thirteen center-of-mass energies 
between 3.97 and 4.26 GeV.  Exclusive cross sections are presented for 
final states consisting of two charmed mesons ($D\bar{D}$, 
$D^{*}\bar{D}$, $D^{*}\bar{D}^{*}$, $D_s^+ D_s^-$, $D_s^{*+} D_s^-$, 
and $D_s^{*+} D_s^{*-}$) and for processes in which the
charmed meson pair is accompanied by a pion.
\end{abstract}

\maketitle

\thispagestyle{fancy}
\section{Introduction}
Hadron production in electron-positron annihilations just above $c{\bar c}$
threshold has been a subject of mystery and little intensive study for
more than three decades since the discovery of charm.  Recent developments,
like the observation of the $Y(4260)$ reported by the BaBar collaboration 
\cite{Aubert:2005rm} and subsequently confirmed by CLEO-c \cite{Coan:2006rv}
and Belle \cite{Abe:2006hf}, underscore 
our incomplete understanding and demonstrate the 
potential for discovery of new states, such as hybrids and glueballs.  It is also 
clear that precise measurements of charm-meson properties are essential for 
higher-energy investigations of $b$-flavored particles and new states that might 
decay into $b$.  They also offer unique opportunities to test the validity and 
guide the development of theoretical tools, like Lattice QCD, that are needed
to interpret measurements of the CKM quark-mixing parameters \cite{Kobayashi:1973fv}.  
Any comprehensive program of precise charm-decay measurements demands a detailed 
understanding of charm production.

Past studies of hadron production in the charm-threshold region have been dominated 
by measurements of the cross-section ratio 
$R=\sigma(e^+e^- \rightarrow {\mathrm{hadrons}}, s)/\sigma(e^+e^- \rightarrow \mu^+ \mu^-, s)$
that have been made over this energy range by many experiments \cite{Eidelman:2004wy}.
Recent measurements with the Beijing Spectrometer (BES) \cite{Bai:2001ct} near 
charm threshold are especially noteworthy. There is a rich structure in 
this energy region, reflecting the production of $c {\bar c}$ resonances and the crossing
of thresholds for specific charm-meson final states. Interesting features in the hadronic 
cross section between 3.9 and 4.2~GeV include a large enhancement at the threshold for 
$D^*\bar{D}^*$ production ($\sim4$~GeV) and a fairly large plateau that begins at  
$D_{s}^{*+}D_{s}^{-}$ threshold. While there is considerable theoretical 
interest \cite{Barnes:2004fs, Barnes:2004cz, Eichten:1979ms, Voloshin:2006pz}, 
there has been little experimental information about the composition of 
these enhancements.

In this paper we describe measurements of charm-meson production in $e^+e^-$
annihilations at thirteen center-of-mass energies between 3970 and 4260~MeV.  These
studies were carried out with the CLEO-c detector at the Cornell Electron
Storage Ring (CESR) \cite{Briere:2001rn} in 2005-6.  (Throughout this paper charge-conjugate 
modes are implied.)  The principal objective of the CLEO-c energy scan was to determine 
the optimal running point for studies of $D_{s}^+$-meson decays.  The same data sample 
has been used to confirm the direct production of $Y(4260)$ in $e^+e^-$ annihilations 
and to demonstrate $Y(4260)$ decays to final states in addition to $\pi^{+}\pi^{-}{\rm{J}}/\psi$ \cite{Coan:2006rv}
. 
Specific results presented in this paper include cross-section measurements for 
exclusive final states with $D^+$, $D^0$ and $D_s^+$ mesons and inclusive measurements 
of the total charm-production cross section and $R$.

\section{Data Sample and Detector}
The data sample for this analysis was collected with the CLEO-c detector.  
Both the fast-feedback analysis carried out as data were collected and the detailed 
analysis reported here are extensions of techniques developed for charm-meson studies 
at the $\psi(3770)$ \cite{He:2005bs}.

An initial energy scan, conducted during August-October, 2005, consisted of twelve 
energy points between 3970 and 4260~MeV, with a total integrated luminosity of 
60.0~pb$^{-1}$.  The scan was designed to provide cross-section measurements at each 
energy for all accessible final states consisting of a pair of charmed mesons.  
At the highest energy point these include $D\bar{D}$, $D^{*}\bar{D}$, $D^{*}\bar{D}^{*}$, 
$D_s^+ D_s^-$, $D_s^{*+} D_s^-$, and $D_s^{*+} D_s^{*-}$, where the first three contain both  
charged and neutral states.  A follow-up run beginning early in 2006 provided a larger 
sample of 178.9 pb$^{-1}$ at 4170~MeV that proved essential in understanding the composition 
of charm production throughout this energy region.  
The center-of-mass energies and integrated luminosities for the thirteen subsamples are 
listed in Table~\ref{tab:Samples}.
\begin{table}[htpb]
\caption{Center-of-mass energies and integrated luminosity totals for all data
samples for the CLEO-c energy scan.}
\label{tab:Samples}
\begin{center}
\begin{tabular}{|c|c|}
\hline
 $E_{\mathrm{cm}}$ (MeV) & $\int \cal{L}\,$ d$t$ (pb$^{-1}$) \\ 
\hline
$3970 $ & $3.85$ \\
$3990 $ & $3.36$ \\
$4010 $ & $5.63$ \\
$4015 $ & $1.47$ \\
$4030 $ & $3.01$ \\
$4060 $ & $3.29$ \\
$4120 $ & $2.76$ \\
$4140 $ & $4.87$ \\
$4160 $ & $10.16$ \\
$4170 $ & $178.89$ \\
$4180 $ & $5.67$ \\
$4200 $ & $2.81$ \\
$4260 $ & $13.11$ \\
\hline
\end{tabular}\end{center}\end{table}
Integrated luminosity is determined by measuring the processes $e^+e^-\rightarrow{e^+e^-}$, $\mu^+\mu^-$, and $\gamma\gamma$, which are used since their cross sections are precisely determined by QED.  Each of the three final states relies on different components of the detector, with different systematic effects. The three individual results are combined using a weighted average to obtain the luminosity needed for this analysis.
%, with normalization provided by the {\textsc Babayaga} event 
%generator \cite{Babayaga}.

CLEO-c is a general-purpose magnetic spectrometer with most components inherited
from the CLEO III detector \cite{CLEO}, which was designed primarily to study 
$B$ decays at the $\Upsilon(4S)$.  Its cylindrical charged-particle tracking 
system covers 93\% of the full 4$\pi$ solid angle and consists of a six-layer 
all-stereo inner drift chamber and a 47-layer main drift chamber.  These 
chambers are coaxial with a superconducting solenoid that provides a uniform 
1.0-Tesla magnetic field throughout the volume occupied by all active detector 
components used for this analysis.  Charged particles are required to satisfy 
criteria ensuring successful fits and vertices consistent with the $e^+e^-$ 
collision point.  The resulting momentum resolution is $\sim 0.6\%$ at 
1~GeV/$c$.  Oppositely-charged and vertex-constrained pairs of tracks are 
identified as $K_S^0 \rightarrow \pi^+ \pi^-$ candidates if their invariant 
mass is within 4.5 standard deviations ($\sigma$) of the known mass ($\sim 12$~MeV/$c^2$).

The main drift chamber also provides $dE/dx$ measurements for charged-hadron 
identification, complemented by a Ring-Imaging Cherenkov (RICH) detector covering 
80\% of 4$\pi$.  The overall efficiency for pion or kaon identification is greater 
than 90\%, and the misidentification probability is less than 5\%.  

An electromagnetic calorimeter consisting of 7784 CsI(Tl) crystals provides 
electron identification and neutral detection over 95\% of 4$\pi$, with 
photon-energy resolution of 2.2\% at 1~GeV and 5\% at 100~MeV.  We select $\pi^0$ 
and $\eta$ candidates from pairs of photons with invariant masses within 3$\sigma$
of the known values \cite{Eidelman:2004wy}($\sim 6$~MeV/$c^2$ for $\pi^0$ and $\sim 12$~MeV/$c^2$ 
for $\eta$).  
%Kinematic mass constraints are imposed in subsequent particle
%reconstruction.

\section{Event-Selection Procedures}
\label{sec:evtsel}
The procedures and specific criteria for the selection of $D^+$, $D^0$ and $D_s^+$ 
mesons closely follow previous CLEO-c analyzes and are described in 
Refs.~\cite{He:2005bs} and \cite{Adam:2006me}.  Candidates are identified based on 
their invariant masses and total energies, with selection criteria optimized on a 
mode-by-mode basis.  We use only the cleanest final states for $D^0$ ($K^- \pi^+$) 
and $D^+$ ($K^- \pi^+ \pi^+$) selection, since these provide sufficient statistics 
for precise cross-section determinations.  For $D_s^+$ 
we optimize for efficiency by selecting eight decay modes: $\phi \pi^+$, $K^{*0} K^+$,
$\eta \pi^+$, $\eta \rho^+$, $\eta' \pi^+$, $\eta' \rho^+$, $\phi \rho^+$, and
$K_S^0 K^+$.  Accepted intermediate-particle decay modes (mass cuts) are 
$\phi \rightarrow K^+ K^-$ ($\pm10$~MeV), $K^{*0} \rightarrow K^- \pi^-$ ($\pm 75$~MeV), 
$\eta' \rightarrow \eta \pi^+ \pi^-$ ($\pm 10$~MeV), and $\rho^+ \rightarrow \pi^+ \pi^0$
($\pm 150$~MeV).

\begin{figure}
\includegraphics[bb= 0 0 551 414, width=80mm]{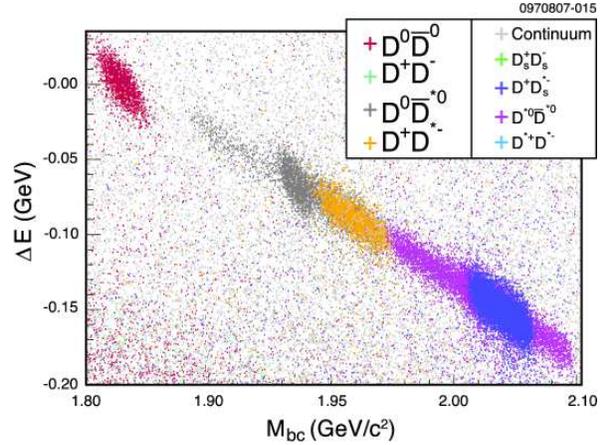}
\caption{$\Delta E$ vs. $M_{\rm bc}$ in a Monte Carlo simulation of expected
final states at a center-of-mass energy of 4160~MeV, showing clear separation
among the expected two-charm-meson final states.}
\label{fig:MbcDE_4160MC}
\end{figure}

\begin{figure}
\includegraphics[bb= 0 0 511 379, width=80mm]{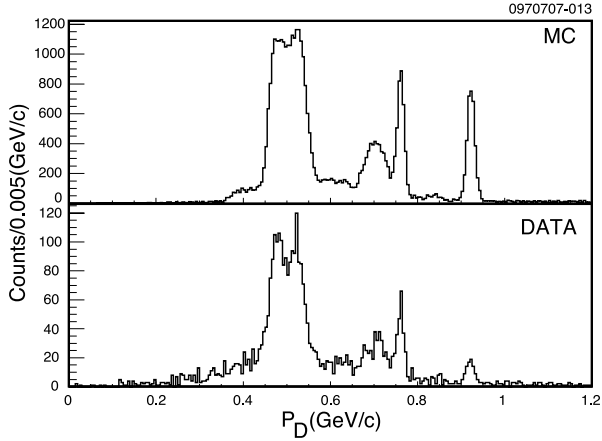}
\caption{Momentum spectra at 4160~MeV for $D^0 \rightarrow K^- \pi^+$
candidates with invariant masses within 15~MeV of the nominal value for
Monte Carlo (top) and data (bottom).  As described in the text, the concentrations 
of entries correspond to the dominant expected final states with two charmed mesons.}
\label{fig:D0_kpi_mom_4160dataMC}
\end{figure}

To determine the production yields and cross sections for the final states accessible
at a particular center-of-mass energy, we classify events based on the energy of a 
$D_{(s)}$ candidate ($\Delta E \equiv E_{\rm beam} - E_{D_{(s)}}$) and
its momentum in the form of beam-constrained mass 
($M_{\rm bc} \equiv \sqrt{E_{\rm beam}^2 - |\vec{P}_{D_{(s)}}|^2}$).  Figure~\ref{fig:MbcDE_4160MC}
shows the expected behavior in a two-dimensional plot of $\Delta E$ vs. $M_{\rm bc}$
for a Monte Carlo simulation of CLEO-c data at 4160~MeV with about ten
times the statistics of our data sample at that energy.  There is clear separation 
of events into the expected final states consisting of two charmed mesons.  This separation was exploited during the scan run for a fast-feedback "cut-and-count" determination of event yields.  
%This
%separation was exploited for the fast-feedback cross-section measurement carried out
%during the scan running through a cut-and-count determination of event yields. 
It is also evident in plots of the momenta of 
charm-meson candidates selected by cutting on candidate mass that the composition 
of final states can be analyzed by fitting the momentum spectra of $D^0$, $D^+$ and 
$D_s^+$ candidates.  Figure~\ref{fig:D0_kpi_mom_4160dataMC} illustrates this with 
the momentum spectra for $D^0 \rightarrow K^- \pi^+$ candidates within 15~MeV of the 
nominal mass both in the Monte Carlo sample of Fig.~\ref{fig:MbcDE_4160MC} and in 
10.16~pb$^{-1}$ of CLEO-c data at 4160~MeV.  While no background corrections have been 
applied to these distributions, the structure of distinct Doppler-smeared peaks 
corresponding to different final states is evident.  The Monte Carlo and data show good 
qualitative agreement, with concentrations of events corresponding to the dominant final 
states near 0.95~GeV/$c$ ($D \bar{D}$), 0.73~GeV/$c$ ($D^* \bar{D}$) and 0.5~GeV/$c$ 
($D^* \bar{D}^*$).
The cross sections for all contributing final states can be determined by correcting 
the raw measured momentum spectra like Fig.~\ref{fig:D0_kpi_mom_4160dataMC} for 
combinatoric and other backgrounds and then fitting to Monte Carlo predictions of 
the spectra.  To achieve good fits, all significant production mechanisms must be 
included and the spectra predicted by Monte Carlo must reflect correct $D^*$-decay angular 
distributions and the effects of initial state radiation (ISR).

\section{Evidence for Multi-Body Production}
While the qualitative features of the measured charm-meson momentum spectra 
accorded with expectations (Fig.~\ref{fig:D0_kpi_mom_4160dataMC}), initial attempts 
to fit the spectra did not produce acceptable results.  It was quickly concluded that 
the two-body processes listed above are insufficient to account for all observed 
charm-meson production.  Final states like ${D\bar{D}^{(*)}{\mathrm{n}}\pi}$, in which the charm-meson pair is accompanied by one or more
additional pions, emerged as the likely explanation.  While not unexpected, these ''multi-body" events have not previously been observed in the charm-threshold
region, and there 
are no predictions of the cross sections for $D^0$ and $D^+$ production 
through multi-body final states.

\begin{figure}
\includegraphics[bb=0 0 492 574, width=80mm]{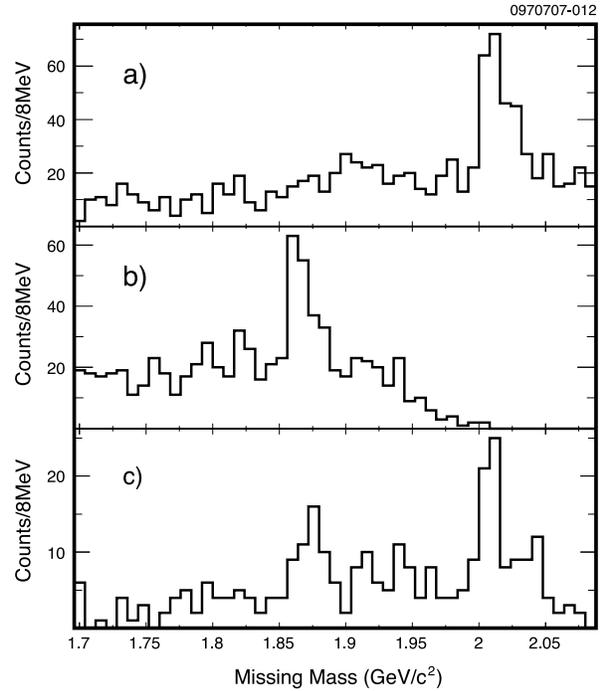}
\caption{The mass spectrum of $X$ in (a) $e^+e^-\rightarrow{D^{0}\pi^{\pm}X}$
at 4170~MeV, (b) $e^+e^-\rightarrow{D^{*\pm}\pi^{\mp}X}$ at 4170~MeV, and
(c) $e^+e^-\rightarrow{D^{*0}\pi^{\pm}X}$ at 4260~MeV.  Peaks at the $D^{*}$ 
mass in (a) and the $D$ mass in (b) are evidence for the decay $D^{*}\bar{D}\pi$.  The 
$D$ peak in (c) confirms $D^*\bar{D}\pi$ and the $D^*$ peak demonstrates that 
$D^*\bar{D}^*\pi$ is produced at 4260 MeV.}
\label{fig:MultiBody_DSDPiPlus_4170}
\end{figure}

To assess which multi-body final states ($D \bar{D} \pi$, $D^{*} \bar{D} \pi$, 
etc.) are measurably populated in our data we examine 
observables other than the charm-meson momenta, because ISR causes smearing 
of the peaks in the momentum spectra that can obscure the two-body kinematics.  We applied $D^{(*)}$ momentum cuts to 
exclude two-body contributions and examined the distributions of missing mass 
against a $D^{(*)}$ and an accompanying charged or neutral pion, using charge 
correlations to suppress incorrect combinations.  Figure~\ref{fig:MultiBody_DSDPiPlus_4170} 
shows clear evidence for $D^{*}\bar{D}\pi$ events at 4170~MeV,
as well as indications of $D^{*}\bar{D}^*\pi$ in the sample of $13~{\rm{pb}}^{-1}$  
collected at 4260~MeV (Fig.~\ref{fig:MultiBody_DSDPiPlus_4170}c). 
These events cannot be attributed to two-body production with ISR, because 
radiative photons would destroy any peaking in the missing-mass 
spectrum.  The absence of a peak at the $D$ mass in 
Fig.~\ref{fig:MultiBody_DSDPiPlus_4170}a indicates that there is no evidence 
for $D\bar{D}\pi$ production.  Analysis of events with $D_{s}$ reveals no 
evidence for multi-body production, consistent with expectations, since the 
$D_s^+ D_s^- \pi^0$ final state violates isospin conservation. 

\section{Momentum-Spectrum Fits and Cross-Section Results}
Candidate momentum spectra for $D^0$, $D^+$ and $D_s^+$ were selected by requiring the invariant
mass to be within $\pm 15$~MeV of the nominal value.  Backgrounds are 
estimated with a sideband technique.  Sideband regions are taken on both sides of 
the expected signal, and are significantly larger than the signal region to 
minimize statistical uncertainty in the background subtraction.  Specific widths 
are set mode by mode based on the expectation of specific background processes. 

Having identified the components of multi-body charm production, we determine
yields for these channels and the two-body modes by fitting the sideband-subtracted 
$D^0$, $D^+$ and $D_s^+$ momentum spectra.  Signal momentum distributions for
specific channels are based on full {\tt GEANT} \cite{GEANT} simulations 
using {\tt EvtGen} \cite{Lange:2001uf} for the production and decay of charmed 
mesons.  The {\tt EvtGen} simulation 
incorporates all angular and time-dependent correlations by using individual 
amplitudes for each node in the decay chain.  ISR is included in the
simulation, which requires input of energy-dependent cross sections for each final 
state.  We used simple parameterizations of these cross sections constructed by
linearly interpolating between the preliminary measurements from our analysis.
(In doing this we make the assumption that the energy dependence
of the Born-level cross sections is adequately represented by the uncorrected
cross sections.)  For the multi-body $D^{*}\bar{D}\pi$
and $D^*\bar{D}^*\pi$ final states we used a spin-averaged phase-space model 
within {\tt EvtGen}.  

\begin{figure}
\includegraphics[bb=0 0 513 748, width=80mm]{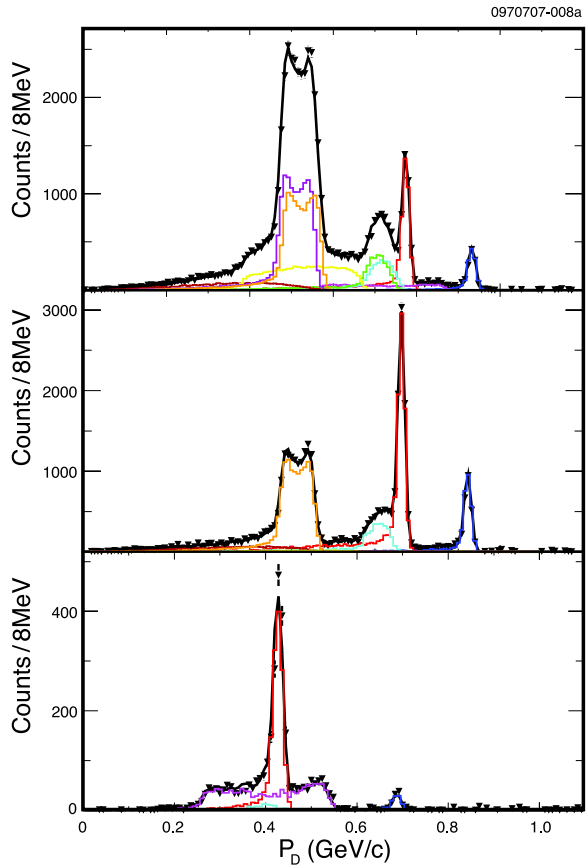}
\caption{Sideband-subtracted momentum spectra for (a)
$D^0\rightarrow{K^-}\pi^+$, (b) $D^+\rightarrow{K^-}\pi^+\pi^+$, 
and (c) $D_{s}^+\rightarrow\phi\pi^+$ (bottom) at 4170~MeV.  Data are shown as points with errors and the total fit result is shown as the solid black line.  The colored histograms represent fit components, mostly single
$D$-production modes.  For example, the primary $D^0$ in $D^{*0}\bar{D}^{*0}$, which peaks at 0.7 GeV/$c$, is shown in bright red.  The secondary  $D^{0}$ mesons from the primary $D^{*0}$
decaying via the emission of a $\pi^{0}$ form the broad peak at 0.6 GeV/$c$ shown in light blue.  The second broad peak, at 0.6 MeV/$c$, consists of $D^0$ mesons from the charged pion decay of the $D^{*+}$ in $D^{*+}D^-$.   All sources of multi-body events are combined and result in the broad spectrum between 0 and 0.5 GeV/$c$ shown in dark red.}
\label{fig:Combined_fit_results_4170}
\end{figure}

Momentum fits for the large sample of data at 4170~MeV are shown in 
Fig.~\ref{fig:Combined_fit_results_4170} for (a) $D^0\rightarrow{K^-}\pi^+$,
(b) $D^+\rightarrow{K^-}\pi^+\pi^+$, and (c) $D_{s}^+\rightarrow\phi\pi^+$ 
candidates.  The lack of $D_s^+$ entries 
below $\sim 200$ MeV confirms the absence of multi-body $D_{s}$ production.  
Because of the relative simplicity of $D_s$ production demonstrated by
the $D_{s}^+\rightarrow\phi\pi^+$ fits and the limited statistics in the sample,
we determined the final cross sections for $D_s^+ D_s^-$, $D_s^{*+} D_s^-$, 
and $D_s^{*+} D_s^{*-}$ by using a sideband subtraction technique to count signal events in a region of the $M_{\rm bc}$ and $\Delta E$ plane.  The
cross sections are then determined from a weighted sum of the yields for the 
eight $D_{s}$ decay modes listed in Sect.~\ref{sec:evtsel}, with weights minimizing the combined statistical and systematic uncertainties that were calculated from previously measured branching fractions and efficiencies determined by Monte Carlo.   The cut-and-count analysis gives results that are consistent 
with momentum fits.  There is good agreement among the
separately-calculated cross sections for the different $D_s$ decay modes.

\begin{figure}
\includegraphics[bb=0 0 498 674, width=80mm]{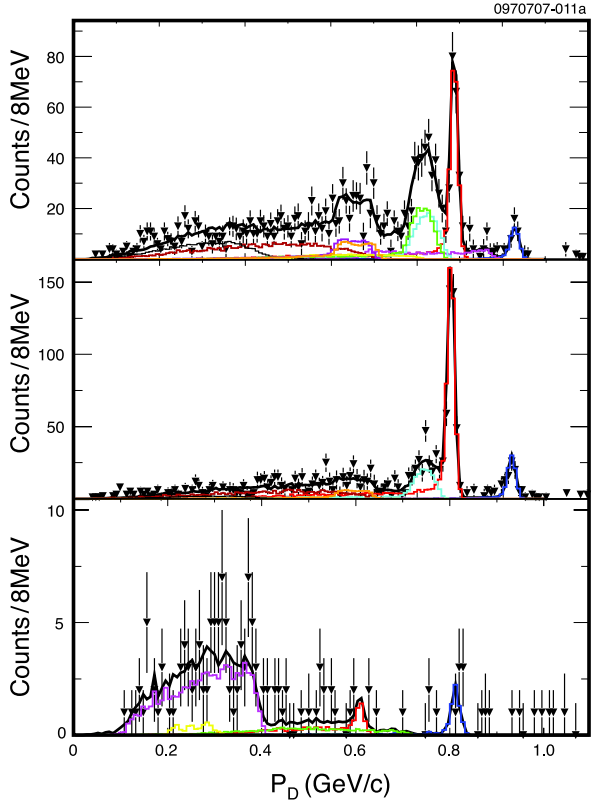}
\caption{Sideband-subtracted momentum spectra for (a)
$D^0\rightarrow{K^-}\pi^+$, (b) $D^+\rightarrow{K^-}\pi^+\pi^+$, 
and (c) $D_{s}^+\rightarrow\phi\pi^+$ (bottom) at 4260~MeV. 
Data are shown as points with errors and the total fit result is shown as the solid black line.  The colored histograms represent fit components, mostly single $D$-production modes.  For example, the primary $D^0$ in $D^{*0}\bar{D}^{*0}$, which peaks at 0.8 GeV/$c$, is shown in bright red.  The secondary  $D^{0}$ mesons from the primary $D^{*0}$ decaying via the emission of a $\pi^{0}$ form the broad peak at 0.7 GeV/$c$ shown in light blue.  The second broad peak, at 0.7 MeV/$c$, consists $D^0$ mesons from the charged pion decay of the $D^{*+}$ in $D^{*+}D^-$.   The multi-body events are combined and result in the broad spectrum between 0 and 0.6 GeV/$c$ shown in dark red for $D^*\bar{D}\pi$ and  in black between 0 and 0.4 GeV/$c$ for $D^*\bar{D}^*\pi$.}
\label{fig:Combined_fit_results_4260}
\end{figure}

Each of the thirteen data subsamples has been analyzed with the techniques 
developed and refined on data at 4170~MeV.  A complete set of fit results is 
provided in Ref.~\cite{Lang_thesis}. Figure~\ref{fig:Combined_fit_results_4260} 
shows the $D^0$, $D^+$ and $D_s$ fits for data sample at 4260~MeV, which are of
particular interest because the charm-production cross sections might 
provide insight to the nature of the $Y(4260)$ state.  The fits at 4260~MeV 
behave similarly to those at lower energy, although a larger proportion of multi-body decays is apparent.

\begin{figure}
\includegraphics[bb= 0 0 501 759, width=80mm]{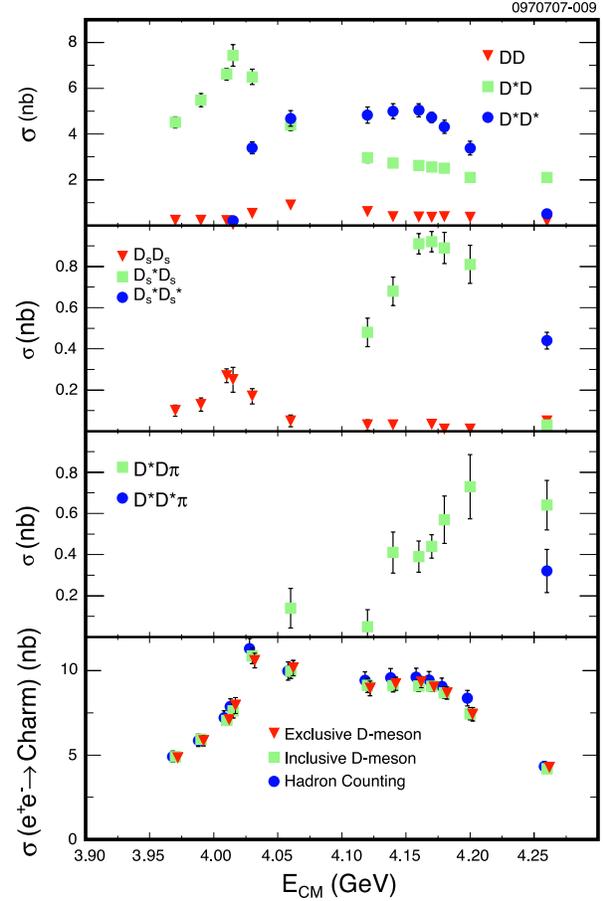}
\caption{Exclusive cross sections for two-body and multi-body 
charm-meson final states, and total observed charm cross section with
combined statistical and systematic uncertainties.}
\label{FIG:Concl_sysDD}
\end{figure}

Cross sections for the two-body and multi-body final states are shown in 
Fig.~\ref{FIG:Concl_sysDD}a-c.  The uncertainties
on the data points are statistical and systematic combined in quadrature.  
Ref.~\cite{Lang_thesis} provides detailed descriptions of the systematic 
uncertainties of the cross-section determinations.  
Briefly, there are three sources of systematic uncertainty: determination 
of the efficiency of charm-meson selection,
extraction of yields, and overall normalization.  The total systematic
uncertainty is not dominated by any one of these.

Track selection and particle identification closely follow previous CLEO-c analyses \cite{He:2005bs, Adam:2006me}.  The efficiency for reconstructing
charged tracks has been estimated by a missing-mass technique applied to events
collected at the $\psi(2S)$ and $\psi(3770)$ resonances.  There is good agreement
between data and Monte Carlo, with an estimated relative uncertainty of $\pm 0.7\%$ per
track.  Pion and kaon identification have been studied with $D^0$ and $D^+$ decays in   
$\psi(3770)$ data, with estimated systematic uncertainties in the respective
efficiencies of $\pm 0.3\%$ and $\pm 1.3\%$.

The extraction of event yields by fitting the charm-meson momentum spectra 
(non-$D_s$ modes) incurs systematic uncertainty primarily through the signal 
functions generated by Monte Carlo, which depend on details of ISR and, in the 
case of $D^*\bar{D}^*$, the helicity amplitudes \cite{Lang_thesis} and resulting $D$-meson angular  
distributions.  As for the exclusive measurements, these details were studied 
with the large data sample at 4170~MeV, for which statistical uncertainties are 
small, and the resulting estimated relative systematic uncertainties are applied 
to all energy points.  For the ISR calculation, the exclusive cross 
sections input to {\tt EvtGen} were varied from their nominal shapes.  While a 
qualitative constraint of consistency with our measured cross sections was imposed, 
some extreme variations are included in the final systematic uncertainty.  
Both the direct effect on the fitted yield of varying a specific mode and the 
indirect effect of varying other modes were computed, although the former dominates 
in quadrature.

The yields for $D_s$ final states are determined by direct counts after
cutting on $M_{\rm bc}$ and $\Delta E$.  Systematic uncertainty arises in
these measurements if the Monte Carlo simulation does not provide an accurate 
determination of the associated efficiency.  This is probed by adjusting the 
cuts and recomputing the cross sections, again using the high-statistics sample
at 4170~MeV.  The systematic uncertainties assigned based on these studies
are $\pm 3\%$, $\pm 2.5\%$ and $\pm 5\%$ for $D_s^+ D_s^-$, $D_s^{*+} D_s^-$, 
and $D_s^{*+} D_s^{*-}$, respectively.

In converting the measured yields to cross sections we must correct for the
branching fractions of the charm-meson decay modes.  For the non-strange
charmed mesons, only one mode is used and CLEO-c measurements 
\cite{He:2005bs} provide the branching fractions and uncertainties:
$\pm 3.1\%$ for $D^0 \rightarrow K^- \pi^+$ and $\pm 3.9\%$ for
$D^+ \rightarrow K^- \pi^+ \pi^+$.  For $D_s$ modes we use CLEO-c measurements
of the branching fractions for the eight decay modes included in the weighted 
sum \cite{Adam:2006me}.  The world-average value is used for the 
$D^{*+} \rightarrow D^0 \pi^+$ branching fraction, with a systematic uncertainty 
of $\pm 0.7\%$ \cite{Eidelman:2004wy}.  Finally, the cross-section normalization 
also depends on the the absolute determination of the integrated luminosity for 
each data sample, with a systematic uncertainty of $\pm 1 \%$.

A mode-by-mode summary of the systematic uncertainties in the 
exclusive cross-section measurements is provided in Table~\ref{tab:exsystematics}.
\begin{table}
\caption{Total systematic errors in the exclusive cross sections.}
\label{tab:exsystematics}
\begin{tabular}{|c|c|}
\hline
{Mode}&Relative Error $(10^{-2})$
\\ \hline
Determined by Momentum Fits&
\\ \hline
\(D\bar{D}\)&\(4.5\)
\\
\(D\bar{D}^*\)&\(3.4\)
\\
\(D^*\bar{D}^*\)&\(4.7\)
\\
\(D^*\bar{D}\pi\)&\(12.0\)
\\
\(D^*\bar{D}^*\pi\)&\(25.0\)
\\ \hline

Determined by Counting&
\\ \hline
\(D_{s}\bar{D}_{s}\)&\(5.6\)
\\
\(D_{s}\bar{D}_{s}^*\)&\(5.3\)
\\
\(D_{s}^*\bar{D}_{s}^*\)&\(6.8\)
\\ \hline

\end{tabular}
\end{table}

As a cross-check, for the two largest data samples (4170~MeV and 4260~MeV), the multi-body 
cross sections are also determined by fitting the distributions of missing mass against 
detected $D^0 \pi$, $D^+ \pi$ and $D^*\pi$ combinations. While these measurements are less precise, 
they show good agreement with the results of the momentum-spectrum fits.

\section{Inclusive Cross-Section Measurements}
If all final states have been included, the sum of the exclusive 
cross sections should equal the total charm cross section.  We test this supposition
with two inclusive measurements that can also be compared with past 
results.  

The first cross-check is a measurement of the total charm-meson cross section:
\begin{equation}
         \sigma(e^+e^-\rightarrow{D\bar{D}}X)= \frac{\sigma_{D^{0}} +
         \sigma_{D^{+}} + \sigma_{D^{+}_{s}}}{2},
\end{equation}
where the contributing cross sections are defined by 
$\sigma_{D} = {N_{D}}$/${\epsilon{B}\cal{L}}$, where $\epsilon$ and
$B$ are the efficiency and branching fraction for the 
decay mode used ($D^{0}\rightarrow{K^-\pi^+}$, 
$D^{+}\rightarrow{K^-\pi^+\pi^+}$, and 
$D_{s}^{+}\rightarrow{K^-K^+\pi^+}$), $\cal{L}$ is the integrated 
luminosity, and $N_{D}$ is the yield obtained by fitting the 
mass spectrum.  In the case of $D^0$ and $D^+$, the invariant-mass distribution is fitted to a Gaussian signal and 
polynomial background.  For $D_s$, the event-type requirements
are maintained because of the relatively large background for the 
high-yield $K^- K^+ \pi^+$ decay mode.  For our energy points below 
4120~MeV, where $D_s$ production occurs only through $D_s^+D_s^-$,
the yield is extracted by fitting $M_{bc}$ to a Gaussian signal and 
ARGUS background function \cite{Albrecht:1990am}.  For 4120~MeV
and above, event types involving $D_s^{*+}$ contribute.  For all
candidate events that pass the selection requirements for any of
$D_s^+ D_s^-$, $D_s^{*+} D_s^-$, and $D_s^{*+} D_s^{*-}$ (the
last only for 4260~MeV), a fit to the $D_{s}^+$ invariant mass is used to
determine the yield. 

The second cross-check is a determination of the total cross section
made by counting multihadronic charm events.  The contribution of $uds$ 
continuum production is estimated with measurements made at ${\rm{E_{cm}}}=$3671~MeV,
below $c {\bar c}$ threshold, and extrapolated as $1/s$.  Procedures 
for this measurement are identical to those used to determine the 
cross section for 
\(e^+e^-\rightarrow\psi(3770)\rightarrow{\mathrm{hadrons}}\) in CLEO-c data 
at \(E_{\rm{cm}}=3770\)~MeV \cite{Besson:2005hm}.

Figure~\ref{FIG:Concl_sysDD}d shows the inclusive 
measurements (statistical and systematic uncertainties combined in
quadrature) and the sum of the cross sections for the measured exclusive 
final states, without radiative corrections.  The excellent agreement demonstrates that, to current precision, the measured exclusive two- and three-body final states saturate charm production in this region.  Furthermore, charm is demonstrated to account for all production of multihadronic events above the extrapolated uds cross section.

For the inclusive-charm cross-section measurements, the systematic uncertainties 
associated with the per-particle efficiencies for tracking and particle
identification are identical to those of the exclusive measurements.  The
uncertainties in normalization (luminosity and branching fractions) are
also identical.  Systematic uncertainty in the yield extraction
is dominated by the choice of fitting function.  This is evaluated mode
by mode and propagated into overall systematic uncertainties accounting
for all correlations, with combined systematic uncertainties of $\pm 4.3\%$,
$\pm 5.1\%$, and $\pm 8.6\%$ ($\pm 10.6\%$) for $D^0$, $D^+$, and $D_s^+$ below
(above) 4120~MeV.  For the hadron-counting inclusive cross sections, the 
systematic uncertainties are identical to those of 
Ref.~\cite{Besson:2005hm}.

For comparison with other experiments and theory it is necessary
to obtain Born-level cross sections from the observed cross sections 
by correcting for ISR.  We do this by calculating correction factors 
following the method of Kuraev and Fadin \cite{Kuraev:1985hb}, which 
gives the observed cross section at any $\sqrt{s}$:
\begin{equation}
\label{eq:KFEQ}
	\sigma_{\rm{obs}}(s) = \int\limits_{0}^{1} dk \cdot f(k,s)\sigma_{B}(s_{\rm{eff}}),
\end{equation}
\noindent where the Born cross section $\sigma_{B}$ is a function
of the effective center-of-mass energy squared 
($s_{\mathrm{eff}} = s(1-({E_{\gamma}}$/${E_{\rm{beam}}})$)), and $f(k,s)$ is 
the ISR kernel.  The radiative-correction factor is also calculated
following the alternative implementation of Bonneau and Martin
\cite{Bonneau:1971mk}.  A 4\% difference between the two calculations
is taken as the systematic uncertainty in the radiative correction.

\begin{figure}
\centering
\includegraphics[bb= 0 0 537 403, width=80mm]{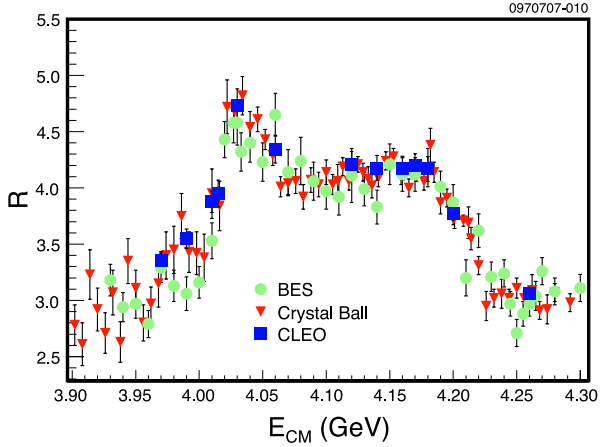}
\caption{$R$ (with radiative corrections as described in the text and correction for non-charm continuum production based on $R_{uds}=2.285\pm0.03$, as determined by a $\frac{1}{s}$ fit to previous $R$ measurements between 3.2 and 3.72 GeV \cite{Osterheld:1986hw}) from this analysis and from
previous measurements \cite{Bai:2001ct,Osterheld:1986hw}.}
\label{RadCor_INR_fig}
\end{figure}

Figure~\ref{RadCor_INR_fig} shows that there is excellent agreement
between our inclusive-charm measurement and $R$ measurements in this region
made by BES \cite{Bai:2001ct} and Crystal Ball \cite{Osterheld:1986hw}.

\section{Summary and Conclusions}
\begin{figure}
\includegraphics[bb=0 0 496 694, width=80mm]{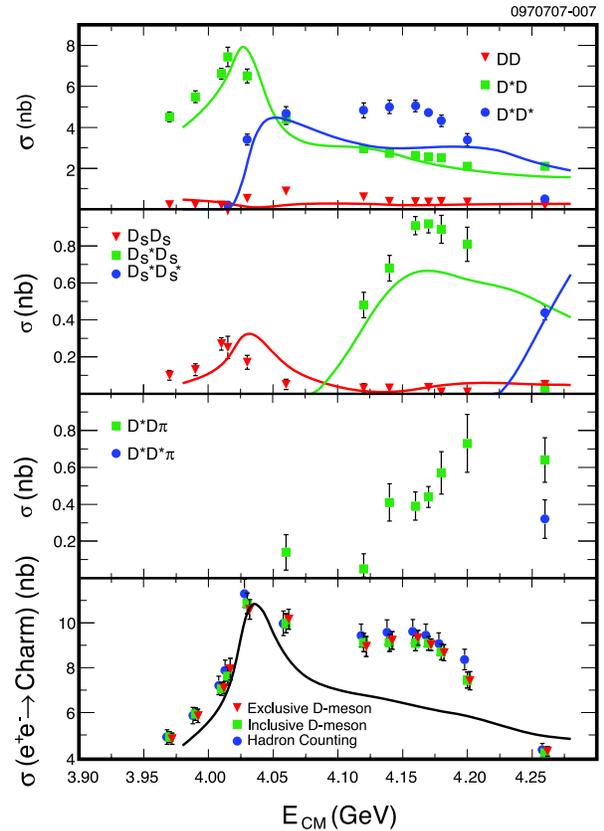}
\caption{Comparisons between measured cross sections and the updated
predictions of the potential model of Eichten {\it et al.} \cite{Eichten:1979ms, Eichten_update} (solid lines).}
\label{fig:Eichten_comparison}
\end{figure}

In summary, we have presented detailed information about 
charm production above $c\bar{c}$ threshold.  Realizing the 
main objective of the CLEO-c scan run, we find the 
center-of-mass energy that maximizes the yield of $D_{s}$ to 
be 4170~MeV, where the cross section of $\sim 0.9$~nb
is dominantly $D_s^{*+} D_s^-$.  This information has guided the
planning of subsequent CLEO-c running, with initial results
already presented on leptonic \cite{Artuso:2006kz} and 
hadronic \cite{Adam:2006me} $D_s$ decays.
The total charm cross section between $3.97$ GeV and $4.26$ GeV 
has been measured both inclusively and for specific two-body
and multi-body final states.  Internal consistency 
is excellent and radiatively-corrected inclusive cross sections are 
consistent with previous experimental results.  
Figure~\ref{FIG:Concl_sysDD} shows that the observed exclusive cross 
sections for $D\bar{D}$, $D^{*}\bar{D}$, $D^{*}\bar{D}^{*}$, 
$D_s^+ D_s^-$, $D_s^{*+} D_s^-$, $D_s^{*+} D_s^{*-}$, 
$D^*\bar{D}\pi$, and $D^*\bar{D}^*\pi$ exhibit structure that 
reflects the intricate behavior expected in the charm-threshold 
region.  Figure~\ref{fig:Eichten_comparison} provides a comparison 
between our measured cross sections and the updated calculation of 
Eichten {\it et al.} \cite{Eichten:1979ms, Eichten_update}.
There is reasonable qualitative 
agreement for most of the two-charm-meson final states.   The most 
notable exception is the cross section for $D^{*}\bar{D}^{*}$ in the 
region between 4050 and 4200 MeV, where the measurement exceeds the 
prediction by as much as 2 nb.  This corresponds to nearly a factor-of-two 
disagreement in the ratio of $D^{*}\bar{D}^{*}$ to $D^{*}\bar{D}$ production,
accounting for about two thirds of the difference in the total charm cross 
section.  This is a much larger effect than the absence of a multi-body 
component from the theoretical prediction.

It has been suggested by Dubynskiy and Voloshin \cite{Dubynskiy:2006sg} that 
the existence of a peak in the 
$D^{*}\bar{D}$ and $D_{s}\bar{D}_{s}$ channels at the $D^{*}\bar{D}^*$ 
threshold, along with the observation that there is a minimum in 
$D\bar{D}$, can be interpreted as a possible new narrow resonance,
but available data are insufficient for a definitive assessment.

The $D^{*}\bar{D}^{*}$ cross section exhibits a plateau just above 
its threshold.  This contrasts with $D^*\bar{D}$, which we observe
to peak at threshold, in agreement with recently presented preliminary 
results from Belle \cite{Abe:2006fj}.  

\begin{acknowledgments}
We gratefully acknowledge the effort of the CESR staff in providing us 
with excellent luminosity and running conditions.  We thank E.~Eichten and
M.~Voloshin for useful discussions. In addition, we thank the organizers of the conference.
\end{acknowledgments}

\bigskip % extra skip inserted

\end{document}